# Nodal-chain network, intersecting nodal rings and triple points coexisting in nonsymmorphic $Ba_3Si_4$

Jin Cai, [a] Yuee Xie, [*a] Po-yao Chang, [b] Heung-Sik Kim, [b] Yuanping Chen[*a]

[a] *School of Physics and Optoelectronics, Xiangtan University, Xiangtan, Hunan, 411105, China*
[b] *Department of Physics and Astronomy, Rutgers University, Piscataway, New Jersey 08854-8019, USA*

*E-mail: xieyech@xtu.edu.cn; chenyp@xtu.edu.cn.

## Abstract

Coexistence of topological elements in a topological metal/semimetal (TM) has gradually attracted attentions. However, the non-topological factors always mess up the Fermi surface and cover interesting topological properties. Here, we find that $Ba_3Si_4$ is a "clean" TM in which coexists nodal-chain network, intersecting nodal rings (INRs) and triple points, in the absence of spin-orbit coupling (SOC). Moreover, the nodal rings in the topological phase exhibit diverse types: from type-I, type-II to type-III rings according to band dispersions. All the topological elements are generated by crossings of three energy bands, and thus they are correlated rather than mutual independence. When some structural symmetries are eliminated by an external strain, the topological phase evolves into another phase including Hopf link, one-dimensional nodal chain and new INRs.

# Introduction

Topological metal and semimetals (TMs) have been new focus after topological insulators.[1-4] Weyl-point, Dirac-point and nodal-line semimetals are three typical TMs.[5-8] The former two have zero-dimensional (0D) topological elements, which are generated by crossing points of (two or four) conduction and valance bands in the momentum spaces.[9-11] The last one has one-dimensional (1D) topological element, which is generated by band crossing along a line.[12-14] Comparing with the nodal-point semimetals, nodal-line semimetals have more subtypes because a line can deform to many different geometries, such as a ring or a knot.[15-20] If there are two or more than two lines in the momentum space, they will construct rich topological phases, such as intersecting nodal rings (INRs),[21-23] nodal link,[24-26] nodal chain[27-30] and Hopf link.[31-33]

Besides the topological elements induced by two or four energy bands, three-, six- and eight-fold band degeneracies have also been proposed.[34-36] Especially the triple point (i.e., triply degenerate point) and its relative topological phases have attracted much attentions. Triple points not only can isolated exist in the momentum space like Weyl or Dirac point,[35, 37, 38] but also can be linked by nodal lines.[39-41] Those points linked by one straight nodal line may evolve into other topological phases, such as Hopf links and nexus networks,[31, 42-44] with the variation of crystal symmetry.

Topological elements are usually protected by specific symmetries, for example, most nodal lines/rings are related to mirror plane,[13, 44] nodal chain is induced by two glide or mirror planes,[27, 45] while triple points are created by three- or four-fold (screw) rotation axis.[31, 39] When a crystal has higher symmetry including many symmetry elements, some topological elements may coexist on the Fermi level and then lead to a composite topological phase.[46-49] These phases not only exhibit various Fermi surfaces, diverse electron/hole pocket patterns, but also provide a platform to study topological phase transition under the variation of crystal symmetry. In most real TMs with composite topological phases, unfortunately, there often exist trivial energy bands crossing over the Fermi level, which mess up the Fermi surface and cover interesting physical properties from topological elements. Therefore, to explore topological properties of TMs with coexisting topological elements, it is necessary to find "clean" TMs.

In this paper, we find that $Ba_3Si_4$ is a good candidate (see Fig. 1). In the absence of spin-orbit coupling (SOC), there are three topological elements coexisting around the Fermi level: a nodal-chain network, INRs and triple points linked by a straight nodal line. All these elements are generated by crossings between three energy bands. The three topological elements are protected by different crystal symmetries, and thus they will disappear or evolve into new topological elements when the corresponding symmetries are eliminated by an external strain. For example, the triple points evolve into a Hopf link and the nodal-chain network into a 1D nodal chain and an isolated ring. It is also found that, in the presence of SOC, the nodal rings are small gaped while the triple points change to Dirac points.

$Ba_3Si_4$ can be synthesized experimentally as a single phase.[50] Figures 1(a) and 1(b) show side and top views of its atomic structure, respectively. It belongs to the nonsymmorphic space group $P4_2/mnm$ and includes 16 symmetry elements, such as three mirror planes ($z = 0$, $y = \pm x$), two glide operations ($G_x$: ($x, y, z$) → (*1/2-x, 1/2+y, 1/2+z*), $G_y$: ($x, y, z$) → (*1/2+x, 1/2-y, 1/2+z*)) and a four-fold screw rotation axis $C_4$ along $z$ axis (see Table S1). The Si atoms are arranged as butterfly shape (see the blue dots in Figs.1(a) and 1(b)). The optimized lattice constants are $a = b = 8.58$ Å and $c = 11.92$ Å . The first Brillouin zone (BZ) of $Ba_3Si_4$ is shown in Fig. 1(c).

## Computational methods

We performed first-principles calculations within the density functional theory (DFT) as implemented in the VASP codes.[51] The potential of the core electrons and the exchange-correlation interaction between the valence electrons were described, respectively, by the projector augmented wave and the generalized gradient approximation (GGA) with Perdew-Burke-Ernzerhof (PBE) functional.[52] A kinetic energy cutoff of 500 eV was used. The atomic positions were optimized using the conjugate gradient method,[53] and the energy and force convergence criteria were set to be $10^{-5}$ eV and $10^{-3}$ eV/Å, respectively. A 5 × 5 × 4 Monkhorst-Pack grid was employed to sample the momentum space.

## Results and discussion

The calculated band structure for $Ba_3Si_4$ is depicted in Fig. 2(a). When SOC is absent, the system can be considered as a spinless system. Along $\Gamma - Z$, a nondegenerate electron-like band and two hole-like degenerate bands cross each other below the Fermi level, and thus the crossing point is a triple point (labeled as $T_1$). Along $\Gamma - X$, $\Gamma - Y$, and $\Gamma - R$, the degeneracy of hole-like bands is broken slightly except at $\Gamma$. The magnified plot in Fig. 2(b) shows that, along $X - \Gamma - Y$, the three bands cross at four points denoted as $D_1$, $D_2$, $D_3$ and $D_4$. However, along $\Gamma - R$ there is only one crossing point $D_5$ by two bands (see Fig. 2(c)).

To obtain topologically electronic structures in $Ba_3Si_4$, we search the full first BZ of the structure and find that there exist three topological elements: triple points, nodal-chain network and INRs. All the elements are generated by the crossings of three bands. To describe the topological phase more clearly, we label the three bands by numbers 1, 2 and 3, according to their energies (see Figs. 2(b-c)). The three topological elements can be explained by crossings of bands (1, 2) or bands (2, 3). We first consider the topological elements related to bands (1, 2). A contour plot for energy difference ($E_2 - E_1$) of the two bands on the plane $k_z = 0$ is shown in Fig. 3(a), where a nodal ring with a center at Γ appears. Two points $D_1$ and $D_3$ on the ring correspond to those in Fig. 2(b). Figure 3(b) presents contour plot on the planes $k_{x/y} = 0$ (the cases on the two planes are the same). Along $k_z$ axis, two red points are linked by a straight nodal line through the boundary of BZ. The red points correspond to the triple point $T_1$ in Fig. 2(a), while nodal line is induced by the degenerate bands from $T_1 - Z$. Meanwhile, a nodal ring appears on the plane, which is cut to two parts by the boundary of BZ. An examination indicates that the two mutually vertical nodal rings on the planes $k_{x/y} = 0$ and $k_z = 0$ in fact connect each other by points $D_1$ (or $D_3$), and thus all the nodal rings form a nodal-chain network. Figure 3(c) exhibits all the topological elements for bands (1, 2).

Then we consider topological elements related to bands (2, 3). The contour plot in Fig. 3(d) shows that there is also a nodal ring on the plane $k_z = 0$. Its diameter is slightly greater than that in Fig. 3(a), because the two points $D_2$ and $D_4$ on the ring correspond to those in Fig. 2(b). On the planes $k_{x/y} = 0$, the two red triple points on $k_z$ axis are linked by a straight nodal line through Γ, as shown in Fig. 3(e). This nodal line is induced by the degenerate bands from $Γ - T_1$ in Fig. 2(a). Additionally, there are two nodal rings on the planes $k_x = \pm k_y$, and they mutually intersect with the nodal ring on the plane $k_z = 0$ in Fig. 3(d). Hence, INRs is formed. All the topological elements are shown in Fig. 3(f). Seen from Figs. 3(c) and 3(f), the topological elements for bands (1, 2) and (2, 3) share the two triple points. If we merge all the topological elements into a whole topological phase, the nodal-chain network in Fig. 3(c) entangles with the INRs in Fig. 3(f), and a straight nodal line passes through the two triple points (see Fig. S1(a)).

All the three topological elements include nodal rings/lines, but these 1D

rings/lines have different topological characteristics. For example, the straight nodal line is generated by quadratic bands, while the nodal rings (in the nodal-chain network and INRs) are generated by linear bands. The former has a trivial Berry phase $2\pi$,[42] while the latter has a nontrivial Berry phase $\pi$.[25] As to the topological nodal rings, Ref. [18] proposed that they can be classified to three types according to slopes of the two crossing linear bands. We calculate band structures for some points on different nodal rings, as shown in Fig. S2. All the three types of nodal rings are found in the topological phase, for example, nodal rings on the planes $k_z = 0$, $k_x = \pm k_y$ and $k_{x/y} = 0$ belong to type-I, II and III, respectively.

The topological elements in Figs. 3(c) and 3(f) are protected by different symmetries: the triple points linked by a straight nodal line are protected by $C_4$ screw rotation axis and mirror plane $k_z = 0$, the nodal-chain network is protected by mirror plane $k_z = 0$ and two glide planes $k_{x/y} = 0$, while the INRs are protected by three mirror planes $k_z = 0$ and $k_x = \pm k_y$. Note that, both the nodal-chain network and INRs are formed by contacting of nodal rings, while the contacting points should be located on the intersecting line where two mirror or glide planes commute to each other.[25]

If the crystal symmetry is partially broken, a topological phase transition occurs. As an example, we use a 1% tensile strain along $x$ axis to decrease the symmetry of $Ba_3Si_4$. In Fig. 4, the calculated band structures, contour plots for energy differences and topological elements are shown. By comparing Fig. 4 with Figs. 2 and 3, three transitions induced by strain deserve special attentions. One is the triple points linked by a straight nodal line change to a Hopf link. Seen from Fig. 4(a), the original degenerate bands along $\Gamma - Z$ in Fig. 2(a) split and the triple point $T_1$ evolves into two crossing points $D_7$ and $D_8$. The two points respectively locate at two mutually vertical nodal rings: one is on the plane $k_y = 0$ in Fig. 4(g), the other is on the plane $k_x = 0$ in Fig.

4(e). The two mutually vertical rings form a Hopf link in the BZ. The second transition is the two nodal rings on the planes $k_x = \pm k_y$ disappear, which can be proven by Fig. 4(a) where the crossing of bands 2, 3 opens a gap along $\Gamma - R$. The third transition is the nodal-chain network decomposes to a 1D nodal chain and an isolated nodal ring. In detail, the original contacting point $D_3$ on the $k_y$ axis ($\Gamma$ -Y) in Fig. 2(b) evolves into two points $D_3'$ and $D_6$ (see Figs. 4(b), 4(f) and (h)), correspondingly the nodal chain along $k_y$ disconnects to isolated rings, as shown in Fig. 4(d). Meanwhile, in Fig. 4(c), one can note that the bigger nodal ring on the plane $k_z = 0$ intersects with one ring (on the plane $k_x = 0$) in the Hopf link, and thus forms new INRs. A topological phase under strain is shown in Fig. S1(b).

The former two transitions are easy to be understood by the variations of structural symmetries: the first transition is because the $C_4$ screw rotation axis is eliminated by the strain;[31] while the second transition is because the strain eliminates the corresponding mirror planes $k_x = \pm k_y$. However, the third transition seems to be counterintuitive: the strain does not break the two glide planes and the mirror plane $k_z = 0$, however, the nodal-chain network disconnects along $k_y$ axis while another two nodal rings connect instead. Our analysis indicates that this transition is actually originated from the loss of $C_4$ symmetry. It is known that, the condition for intersecting of two mutually vertical rings is the corresponding two-intersecting mirror or glide planes commute with each other.[25] The nodal-chain network in Fig. 3(c) is related to two glide planes $k_{x/y} = 0$ and a mirror plane $k_z = 0$. Figures 5(a) presents schematic energy bands (along $X - \Gamma - Y$) as well as their eigenvalues in the case of no strain. $m_z$, $n_x$ and $n_y$ represent the eigenvalues of the planes $k_z = 0$, $k_y = 0$ and $k_x = 0$ respectively. The red and black bands have opposite eigenvalues $m_z$ and $n_x$ along $k_x$ axis ($\Gamma$-X), while the blue and black bands have opposite eigenvalues $m_z$ and $n_y$ along $k_y$ axis ($\Gamma$-Y). The blue band along $k_x$ axis and red band along $k_y$ axis are related to each other by the $C_4$ screw rotation symmetry, and thus two mutually vertical nodal rings for bands (1, 2) can intersect along both $k_x$ and $k_y$ axes (see the black stars in Fig. 5(a)). In this case, the nodal rings for bands (1, 2) can form nodal chains along both $k_x$ and $k_y$ directions. When the $C_4$ symmetry is broken by strain, the degeneracy at Γ for the red and blue bands is lifted, meanwhile the blue band is pushed above the red band, as shown in Fig. 5(b). This gives rise to the intersecting of nodal rings only occurring for bands (2, 3) along $k_y$ axis, which is also the reason why the two rings form INRs in Fig. 4(c). For bands (1, 2) along $k_y$ axis, however, two separated crossing points indicate that the nodal chain disconnects to isolated rings.

In addition, we calculate surface states of $Ba_3Si_4$ on surface [001], as shown in Fig. S3. Because the strain increases the distance between the two nodal rings on the plane $k_z = 0$ [see Figs. S3(a) and (c)], the drumhead states in the two projected rings are correspondingly separated. We also calculate band structure for $Ba_3Si_4$ after considering the effect of SOC, as shown in Fig. S4. One can find that, all the crossing points on Γ − X, Γ − Y, and Γ − R open a small gap, while the triple points on $k_z$ axis change to Dirac points. However, because the gaps between the original crossing points are small, for example, only 2~6 meV on Γ − X and Γ − Y, and thus the topological elements mentioned above, especially the nodal rings still can be easily observed.

## Conclusions

In conclusion, we have calculated topologically electronic properties of $Ba_3Si_4$ and find that it is a TM. In the absence of SOC, its topological phase consists of three topological elements: nodal-chain network, INRs and triple points linked by a straight nodal line. These topological elements are generated by crossings of three energy bands near the Fermi level. The nodal rings/lines in the topological phase exhibit diverse topological characteristics, for example, type-I, type-II and type-III nodal rings with different band dispersions are all observed in the momentum space. The three topological elements are protected by different crystal symmetries. Under an external strain along *x* axis, the topological phase evolves into another composite phase including Hopf link, 1D nodal chain and new INRs. The evolution process is analyzed by the variation of structural symmetries.

## Conflicts of interest

There are no conflicts to declare.

## Acknowledgements

This work was supported by the National Science Foundation of China (No. 11474243 and 51376005).

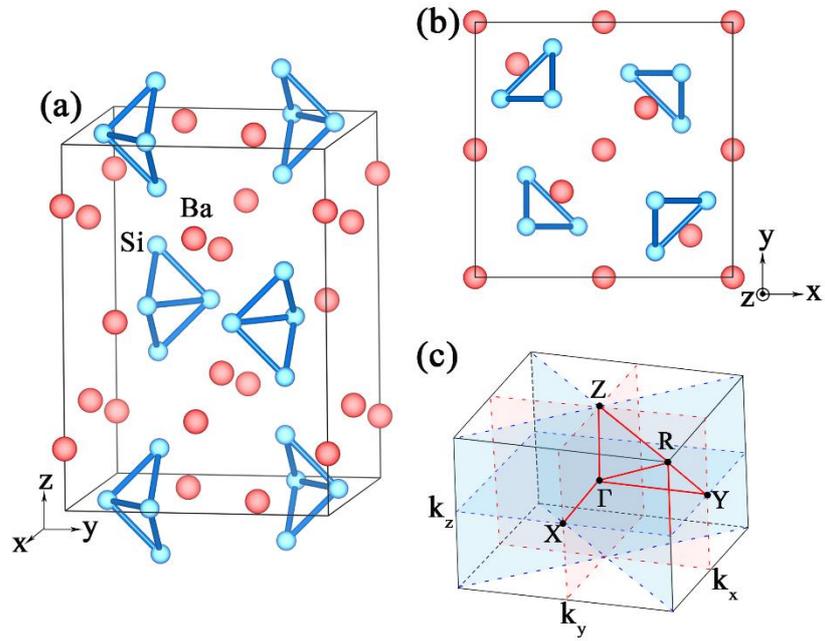

FIG. 1. Side view (a) and top view (b) of the primitive cell of $Ba_3Si_4$, where the blue and red dots represent Si and Ba atoms, respectively. The Si atoms form butterfly shape. (c) BZ of the structure, where the blue and red planes represent mirror and glide planes, respectively.

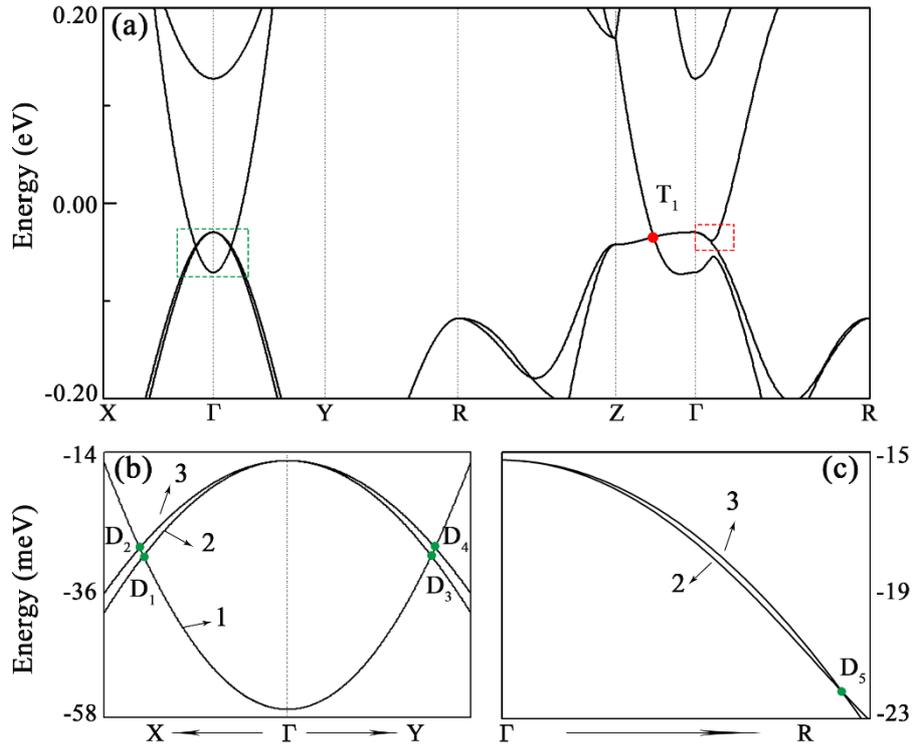

FIG. 2. (a) Calculated band structure of $Ba_3Si_4$ in the absence of SOC, where there are three energy bands around the Fermi level. $T_1$ is a triple point crossed by a degenerate band and a nondegenerate band. (b) and (c) are amplified regions in the green and red dashed boxes in (a), respectively. The numbers 1, 2, 3 represent the three energy bands from low to high energies. $D_1$~$D_4$ in (b) and $D_5$ in (c) are crossing points of bands.

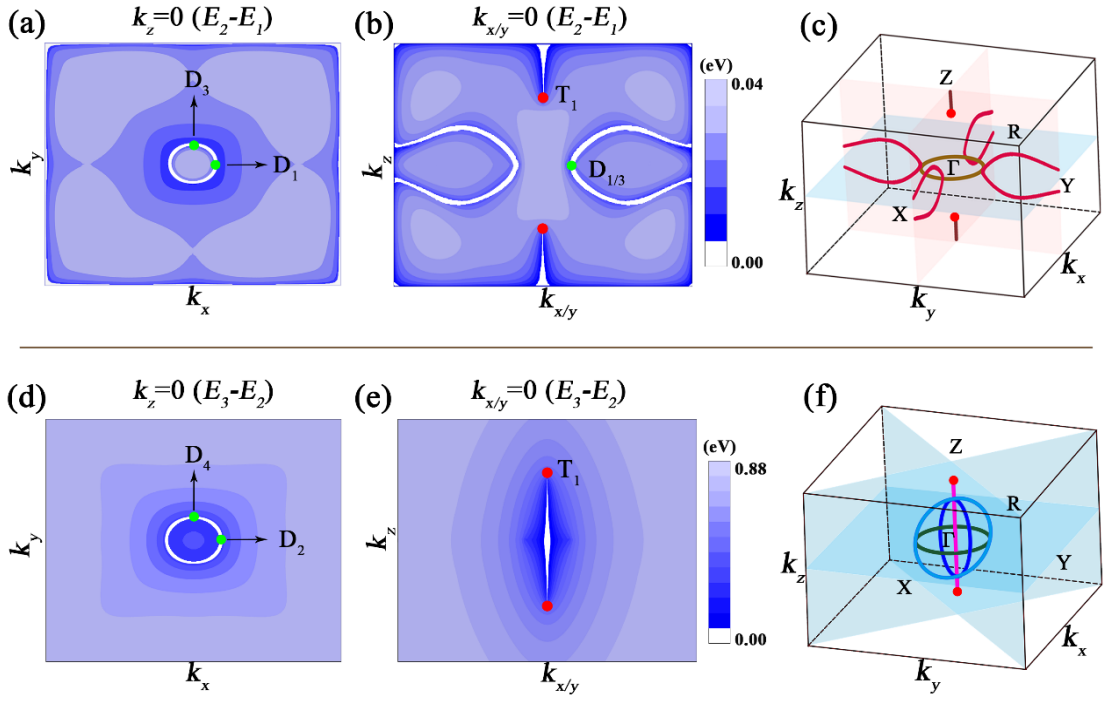

FIG. 3. Contour plots of energy difference $E_2 - E_1$ for bands (1, 2) on the plane $k_z = 0$ (a) and $k_{x/y} = 0$ (b), where $E_2$ and $E_1$ represent the energies of bands 1 and 2 in Figs. 2(b) and 2(c), respectively. (c) Topological elements for bands (1, 2) in the first BZ. Contour plots of energy difference $E_3 - E_2$ for bands (2, 3) on the plane $k_z = 0$ (d) and $k_{x/y} = 0$ (e), where $E_2$ and $E_3$ represent the energies of bands 2 and 3, respectively. (f) Topological elements for bands (2, 3) in the first BZ. $D_1 \sim D_4$ and $T_1$ in (a-b) and (d-e) correspond to the Dirac points and triple point in Fig. 2.

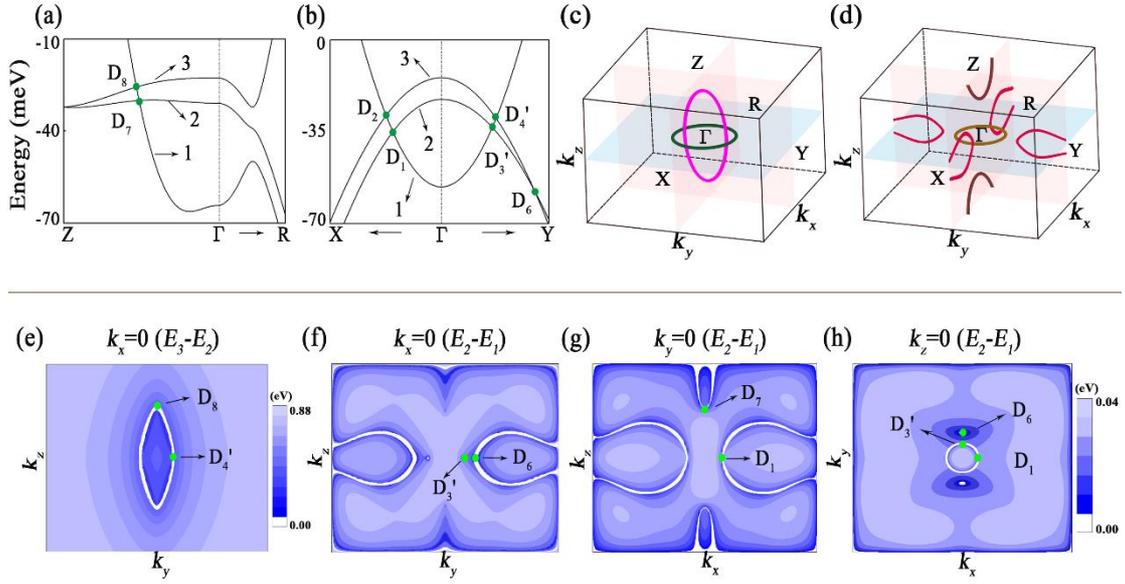

FIG. 4. (a-b) Band structures of $Ba_3Si_4$ under a 1% tensile strain along $x$ axis. Topological elements for bands (1, 2) (c) and (2, 3) (d) in the first BZ. (e-h) Contour plots of energy differences for bands (2, 3) or (1,2) on the mirror or glide planes. $D_7$ and $D_8$ in (a) are crossing points split by the triple points $T_1$ in Fig. 2(a), which is responsible for the triple points with linked straight nodal line splitting to Hopf link.

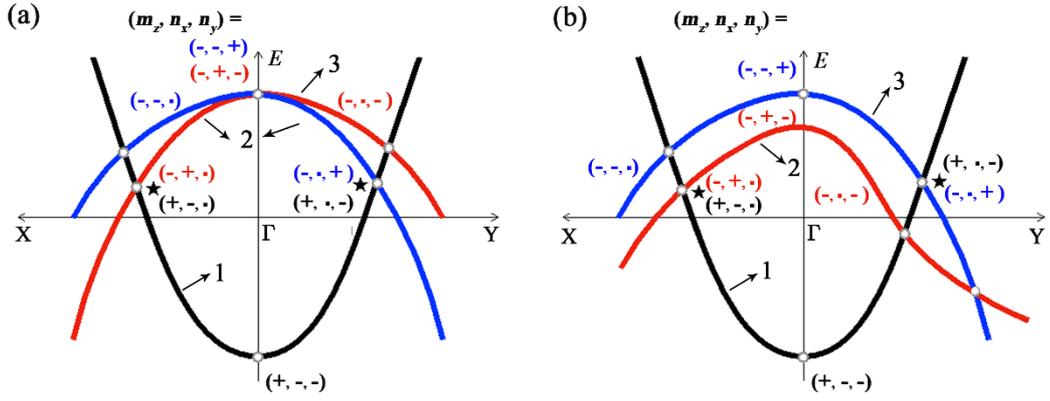

FIG. 5. (a) Schematic band structure along X-Γ-Y with eigenvalues of symmetry in the case of no strain. $m_z$, $n_x$, and $n_y$ represent the eigenvalues of the mirror plane $k_z = 0$, glide planes $k_{y/x} = 0$. All the three symmetries $m_z$, $n_x$, and $n_y$, are preserved at Γ, while $n_x$, and $n_y$ are lost on Γ-Y and Γ-X lines respectively (a dot in the symmetry labels means that the corresponding symmetry is lost). Intersecting points of two mutually vertical rings are marked with black stars. Note that, blue band along Γ-X and red band along Γ-Y are related to each other by the $C_4$ screw rotation symmetry. (b) Schematic band structure in the case of under strain. After the $C_4$ screw rotation symmetry is lost, the blue band is pushed above in energy, opening the nodal intersecting lines into two separated nodal lines on Γ-Y. The numbers 1, 2, and 3 label the three bands according to their energies, same to those in Figs. 2(b-c).